\documentclass[useAMS,usenatbib]{mn2e}
\usepackage{epsfig}
\usepackage{amsmath}

\newcommand{\be}{\begin{equation}}
\newcommand{\beq}{\begin{equation}}
\newcommand{\ba}{\begin{eqnarray}}
\newcommand{\ee}{\end{equation}}
\newcommand{\eeq}{\end{equation}}
\newcommand{\ea}{\end{eqnarray}}

\newcommand{\apj}{ApJ}
\newcommand{\apjl}{ApJL}
\newcommand{\mnras}{MNRAS}
\newcommand{\aj}{AJ}
\newcommand{\apjs}{ApJS}

\def\lsim{~\rlap{$<$}{\lower 1.0ex\hbox{$\sim$}}}

\def\gsim{~\rlap{$>$}{\lower 1.0ex\hbox{$\sim$}}}

\title[21cm observations of Quasar Proximity Zones]{Redshifted 21cm Observations of High Redshift Quasar Proximity Zones}

\author[Wyithe]{J. Stuart B. Wyithe \\
School of Physics, University of Melbourne, Parkville, Victoria,
Australia\\Email: swyithe@unimelb.edu.au}

\begin{document}

%\date{\today}
%\pagerange{\pageref{firstpage}--\pageref{lastpage}} \pubyear{2006}

\maketitle

\label{firstpage}
\begin{abstract}

The introduction of low-frequency radio arrays is expected to
revolutionize the study of the reionization epoch. Observation of the
contrast in redshifted 21cm emission between a large HII region and
the surrounding neutral inter-galactic medium (IGM) will be the
simplest and most easily interpreted signature. However the highest
redshift quasars known are thought to reside in an ionized IGM.  Using
a semi-analytic model we describe the redshifted 21cm signal from the
IGM surrounding quasars discovered using the $i$-drop out technique
(i.e. quasars at $z\sim6$). We argue that while quasars at $z<6.5$
seem to reside in the post overlap IGM, they will still provide
valuable probes of the late stages of the overlap era because the
light-travel time across a quasar proximity zone should be comparable
to the duration of overlap. For redshifted 21cm observations within a
32MHz bandpass, we find that the subtraction of a spectrally smooth
foreground will not remove spectral features due to the proximity
zone. These features could be used to measure the neutral hydrogen
content of the IGM during the late stages of reionization. The density
of quasars at $z\sim6$ is now well constrained. We use the measured
quasar luminosity function to estimate the prospects for discovery of
high redshift quasars in fields that will be observed by the Murchison
Widefield Array.

\end{abstract}


\section{Introduction}

The reionization of cosmic hydrogen by the first stars and quasars
(e.g. Barkana \& Loeb~2001), was an important milestone in the history
of the Universe.  The recent discovery of distant quasars has allowed
detailed absorption studies of the state of the high redshift
intergalactic medium (IGM) at a time when the universe was less than a
billion years old (Fan et al.~2006; White et al.~2003).  Several
studies have used the evolution of the ionizing background inferred
from these spectra to argue that the reionization of cosmic hydrogen
was completed just beyond $z\sim6$ (Fan et al.~2006; Gnedin \&
Fan~2006; White et al.~2003). However, other authors have claimed that
the evidence for this rapid change becomes significantly weaker for a
different choice of density distribution in the IGM (Becker et
al.~2007). Different arguments in favour of a rapidly evolving IGM at
$z>6$ are based on the properties of the putative HII regions inferred
around the highest redshift quasars (Wyithe \& Loeb~2004; Wyithe,
Loeb, \& Carilli 2005; Mesinger \& Haiman 2005). However, Bolton \&
Haehnelt (2007a) and Lidz et al. (2007) have demonstrated that the
interpretation of the spectral features is uncertain and that the
observed spectra could either be produced by an HII region, or by a
classical proximity zone.  One reason for the ambiguity in
interpreting these absorption spectra is that Ly$\alpha$ absorption
can only be used to probe neutral fractions that are smaller than
$10^{-3}$ owing to the large cross-section of the Ly$\alpha$
resonance.  Thus studies of the IGM in Ly$\alpha$ absorption become
inconclusive in the era of interest for reionization.

On the other hand there is mounting evidence that the reionization of
the IGM was photon starved. Firstly Bolton \& Haehnelt~(2007b) have
shown that the observed ionization rate at $z\la6$ implies an
emissivity that is only just sufficient to have reionized the universe
by that time. Similarly, the small escape fractions found for high
redshift galaxies by several studies (Chen et al.~2007; Gnedin et
al.~2007; Srbinovsky \& Wyithe~2008) together with the star formation
rates implied by the observed high redshift galaxy population suggest
a photon budget that struggles to have been sufficient to reionize the
universe by $z\sim6$ (Gnedin~2007; Srbinovsky \& Wyithe~2008). These
results imply that while the IGM seems to be highly ionized along the
lines-of-sight towards the highest redshift quasars discovered in
current surveys, the reionization epoch cannot be at a substantially
higher redshift unless the emissivity grows significantly at
$z>6$. Indeed this outcome may be suggested by the large optical depth
to electron scattering measured by the WMAP satellite (Spergel et
al.~2007).

A better probe of the process of reionization will be provided by
redshifted 21cm observations.  Reionization starts with ionized (HII)
regions around galaxies, which later grow to surround groups of
galaxies. The process of reionization is completed when these HII
regions overlap (defining the so-called {\it overlap} epoch) and
fill-up most of the volume between galaxies.  Several probes of the
reionization epoch in redshifted 21cm emission have been suggested in
a large body of literature. These include observation of 21cm emission
as a function of redshift averaged over a large area of the sky. This
provides a direct probe of the evolution in the neutral fraction of
the IGM, the so-called global step (Shaver, Windhorst, Madau \&
de~Bruyn~1999; Gnedin \& Shaver~2004). A more powerful probe will be
provided by observation of the power-spectrum of fluctuations together
with its evolution with redshift. This observation would trace the
evolution of neutral gas with redshift as well as the topology of the
reionization process (e.g. Tozzi, Madau, Meiksin \& Rees~2000;
Furlanetto, Hernquist \& Zaldarriaga~2004; Loeb \& Zaldarriaga 2004;
Barkana \& Loeb 2005a,b,c). It is thought that the amplitude of 21cm
fluctuations will be greatest when the neutral fraction in the IGM is
around 50\% (Furlanetto et al.~2004; Lidz et al.~2007). Thus, while the
power-spectrum should prove to be the best technique for study of the
bulk of the reionization epoch, it may not be a sensitive probe of the
very late stages of the overlap era.

Finally, observations of individual quasar HII regions will probe
quasar physics as well as the evolution of the neutral gas in the
surrounding IGM (Wyithe \& Loeb~2004b; Kohler, Gnedin, Miralda-Escude
\& Shaver~2005). Kohler et al.~(2005) have generated synthetic spectra
using cosmological simulations. They conclude that quasar HII regions
will provide the most prominent individual cosmological signals.
These individual signatures will be most readily detected
a-posteriori, around known high redshift quasars (Wyithe, Loeb \&
Barnes~2005; Geil \& Wyithe~2007). These studies have focused on the
scenario of a quasar expanding into a significantly neutral
IGM. However the density of quasars is very low at high redshift,
while as discussed above, the IGM allows substantial Ly$\alpha$
transmission and so is thought to be highly ionized along the
lines-of-sight to nearly all of the known high redshift quasars.

The conventional wisdom has been that the 21cm signal disappears after
the {\it overlap} epoch is complete, because there is little neutral
hydrogen left through most of intergalactic space. However
observations of damped Ly$\alpha$ systems out to a redshift of
$z\sim4$ show the cosmological density parameter of HI to be
$\Omega_{\rm HI}\sim10^{-3}$ (Prochaska et al. ~2005). In the standard
cosmological model the density parameter of baryons is $\Omega_{\rm
b}\sim 0.04$, so that the mass-averaged neutral hydrogen fraction at
$z\sim4$ (long after the end of the HII overlap epoch) is $F_{\rm
m}\sim0.03$. This neutral gas does not contribute significantly to the
effective Ly$\alpha$ optical depth, which is sensitive to the volume
averaged neutral fraction (with a value that is orders of magnitude
lower). However the redshifted 21cm emission is sensitive to the total
(mass-weighted) optical depth of this neutral gas.  Observations of
the redshifted 21cm signal would therefore detect the total neutral
hydrogen content in a volume of IGM dictated by the observatory beam
and frequency band-pass (Wyithe \& Loeb~2007). Since quasars could be
observed through the entire overlap epoch, redshifted 21cm
observations of the surrounding IGM could provide a bridge between
21cm fluctuations at high redshift and the well studied techniques
utilizing the Ly$\alpha$ forest following the completion of
reionization.

We begin by describing our density dependent semi-analytic model for
the reionization history (\S~\ref{models}). We next describe our
calculation of the depletion of neutral hydrogen near the vicinity of
a high redshift quasar (\S~\ref{proximity}). Then in
\S~\ref{21cmproximity} and \S~\ref{fgproximity} we describe the 21cm
signal from the proximity zones and estimate of the effect of
foreground removal. Finally we summarize existing observations of the
high redshift quasar luminosity function (\S~\ref{LF}), and predict
the number that will be found in future surveys (\S~\ref{counts})
before presenting our conclusions in \S~\ref{conclusion}. Throughout
the paper we adopt the set of cosmological parameters determined by
{\it WMAP} (Spergel et al. 2007) for a flat $\Lambda$CDM universe,
namely $\Omega_m=0.24$, $\Omega_\Lambda=0.76$ and $h=0.73$. In
computation of the mass function we assume a primordial power spectrum
defined by a power law with index $n=0.95$, an exact transfer function
given by Bardeen et al.~(1986) and rms mass density fluctuations with
a sphere of radius $R_8=8h^{-1}\mathrm{Mpc}$ of $\sigma_8=0.76$.

\section{Semi-Analytic Model for the Reionization History}
\label{models}

\begin{figure*}
\includegraphics[width=15cm]{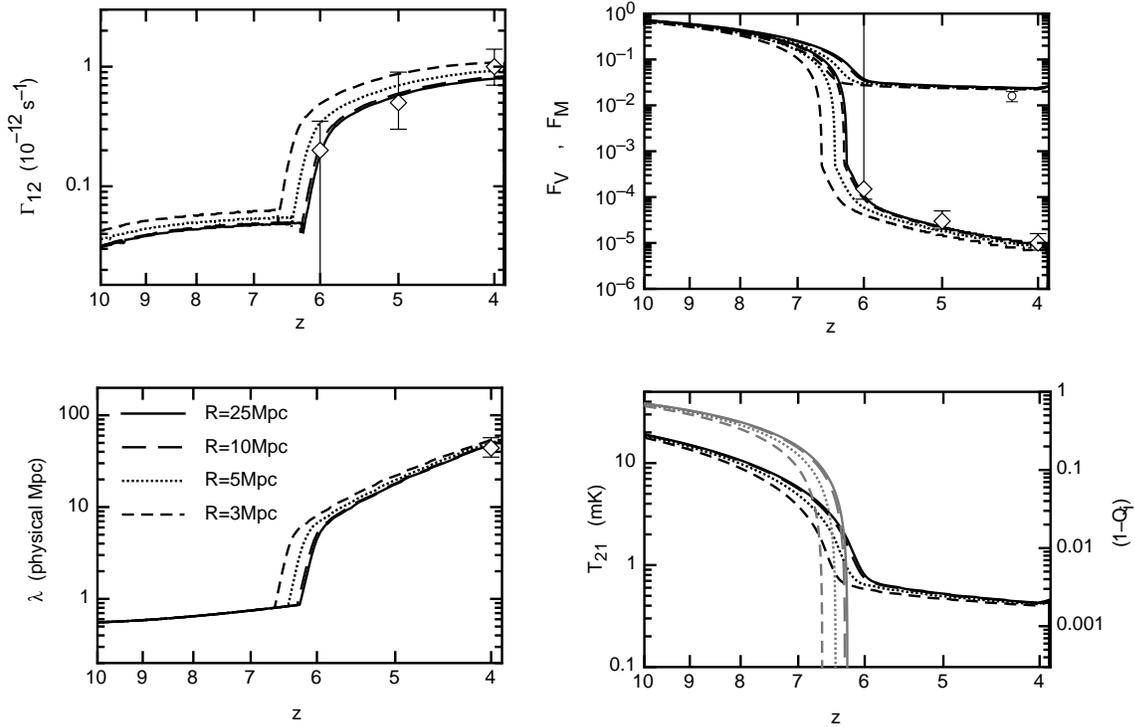} 
\caption{The effect of over density on the redshift of
overlap, and the subsequent ionization state of the IGM. Five cases
are shown, corresponding to over densities evaluated within spheres
with radii of 3, 5, 10 and 25 physical Mpc, centered on a quasar host of
mass $M=10^{13}M_\odot$ at $z\sim6$. In each case we evaluate the
reionization history assuming the mean over density surrounding the
quasar.  {\em Upper Left Panel:} The ionization rate as a function of
redshift. The observational points are from Bolton et
al.~(2007b). {\em Upper Right Panel:} The volume (lower curves) and
mass (upper curves) averaged fractions of neutral gas in the
universe. Also shown (dotted lines) is the fraction of the IGM yet to
overlap $(1-Q_i)$. The observational points for the volume averaged
neutral fraction are from Bolton et al.~(2007b), while the observed
mass-fractions are from the damped Ly$\alpha$ measurements of
Prochaska et al.~(2005).  {\em Lower Left Panel:} The mean-free-path
for ionizing photons computed using the formalism in
\S~\ref{models}. The data points are based on Storrie-Lombardi et
al.~(1994). {\em Lower Right Panel:} The evolution of the mean 21cm
brightness temperature (in mK) with redshift (solid lines). For
comparison, the fraction of IGM yet to overlap $(1-Q_i)$ is over 
plotted.}
\label{fig1}
\end{figure*}

In this section we describe the semi-analytic model which we use to
describe the ionization state of the IGM during the reionization
history of the IGM. Our model is based on the work of Miralda-Escude
et al.~(2000) who presented a prescription which allows the
calculation of an effective recombination rate in an inhomogeneous
universe by assuming a maximum over density ($\Delta_{\rm c}$)
penetrated by ionizing photons within HII regions. The model assumes
that reionization progresses rapidly through islands of lower density
prior to the overlap of individual cosmological ionized
regions. Following the overlap epoch, the remaining regions of high
density are gradually ionized. It is therefore hypothesized that at
any time, regions with gas below some critical over density
$\Delta_{\rm i}\equiv {\rho_{i}}/{\langle\rho\rangle}$ are highly
ionized while regions of higher density are not. In what follows, we
draw primarily from their prescription and refer the reader to the
original paper for a detailed discussion of its motivations and
assumptions.  Wyithe \& Loeb~(2003) employed this prescription within
a semi-analytic model of reionization. This model was extended by
Srbinovsky \& Wyithe~(2007) and by Wyithe, Bolton \& Haehnelt~(2007).
We summarise the model in the remainder of this section, but refer the
reader to those papers for a full description.

Within the model of Miralda-Escude et al.~(2000) we describe the
post-overlap evolution of the IGM by computing the evolution of 
the fraction of mass in regions with over density below $\Delta_{\rm i}$,
\begin{equation}
F_{\rm M}(\Delta_{\rm i})=\int_{0}^{\Delta_{\rm i}}d\Delta P_{\rm
V}(\Delta)\Delta,
\end{equation}
where $P_{\rm V}(\Delta)$ is the volume weighted probability
distribution for $\Delta$.  Miralda-Escude et al.~(2000) quote a
fitting function which provides
 a good fit to the volume weighted probability distribution for the
baryon density in cosmological hydrodynamical simulations. This
probability distribution remains a reasonable description at high
redshift when confronted with a more modern cosmology and updated
simulations, although the addition of an analytical approximation for
the high density tail of the distribution remains necessary as a best
guess at correcting for numerical resolution (Bolton \& Haehnelt~2007b).

In the post overlap era the model computes the evolution of
$\Delta_i$. In the pre-overlap era we define the quantity $Q_{\rm i}$
to be the volume filling factor within which all matter at densities
below $\Delta_{\rm i}$ has been ionized. Within this formalism, the
epoch of overlap is precisely defined as the time when $Q_{\rm i}$
reaches unity. However, prior to overlap we have only a single equation to describe the
evolution of two independent quantities $Q_{\rm i}$ and $F_{\rm M}$
(or equivalently $\Delta_{\rm i}$).  The relative growth of these
depends on the luminosity function and spatial distribution of the
sources. In our model we follow Miralda-Escude et al.~(2000) and
assume $\Delta_{\rm i}$ to be constant\footnote{We note that the
assumption of a fixed value of $\Delta_{\rm c}$ (and hence a slowly
evolving mean-free-path) in the pre overlap era is artificial. Indeed,
$\Delta_{\rm c}$ is probably only a meaningful quantity after overlap
is complete and the mean-free-path is set by dense systems, while
before overlap the mean-free-path is sensitive to the the size of HII
regions (which increases with redshift). However, within the model of
Miralda-Escude et al.~(2000), the value of $\Delta_{\rm c}$ and the
ionization fraction are both unknowns prior to overlap, with only one
equation to govern their evolution. An assumption regarding
$\Delta_{\rm c}$ is therefore unavoidable within the formalism used in
this paper. } (of value $\Delta_{\rm c}$) with redshift before the
overlap epoch, and in this paper compute results for models that
assume $\Delta_{\rm c}=20$. Our approach is to compute a reionization
history given a particular value of $\Delta_{\rm c}$, combined with
assumed values for the efficiency of star-formation and the fraction
of ionizing photons that escape from galaxies. With this history in
place we then compute the evolution of the background radiation field
due to these same sources.  After the overlap epoch, ionizing photons
will experience attenuation due to residual over dense pockets of HI
gas.  We use the description of Miralda-Escude et al.~(2000) to
estimate the ionizing photon mean-free-path, and subsequently derive
the attenuation of ionizing photons. We then compute the flux at the
Lyman-limit in the IGM due to sources immediate to each epoch, in
addition to redshifted contributions from earlier epochs.

We assume the spectral energy distribution (SED) of population-II star
forming galaxies with a gas metalicity of 0.05 and a Scalo IMF, using the model presented in Leitherer et
al.~(1999). The star formation rate per unit volume is computed based
on the collapsed fraction obtained from the extended
Press-Schechter~(1974) model (Bond et al.~1991) in halos above the
minimum halo mass for star formation, together with an assumed star
formation efficiency ($f_\star$).  In a cold neutral IGM beyond the
redshift of reionization, the collapsed fraction should be computed
for halos of sufficient mass to initiate star formation. The critical
virial temperature is set by the temperature ($T_{{\mathrm{N}}}\sim
10^4$ K) above which efficient atomic hydrogen cooling promotes star
formation. Following the reionization of a region, the Jeans mass in
the heated IGM limits accretion to halos above
$T_{{\mathrm{I}}}\sim10^5$ K (Efstathiou~1992; Thoul \& Weinberg~1996;
Dijkstra et al.~2004). Only a
fraction of ionizing photons produced by stars enter the
IGM. Therefore an additional factor of $f_{\mathrm{esc}}$ (the escape
fraction) must be included when computing the emissivity of
galaxies. In our fiducial model we assume this escape fraction to be
independent of mass. We define a parameter $f_{\rm \star,esc}\equiv
f_\star f_{\rm esc}$.

Figure~\ref{fig1} shows models for the reionization of the IGM and the
subsequent post-overlap evolution of the ionizing radiation field. The
fiducial model (shown by the thick grey curves) has $f_{\rm
\star,esc}=0.00375$.  Our model allows the bias of reionization near a
massive halo to be included explicitly, and we show histories
corresponding to regions within $R=3$, 5, 10 and 25 proper Mpc surrounding a
quasar host of mass $M=10^{13}M_\odot$.  In the top left panel of
Figure~\ref{fig1} we show the evolution of the ionization rate. The
observational points are from the simulations of Bolton et al.~(2007b;
based on the observations of Fan et al.~2006).  In the upper-right
panels we plot the corresponding volume and mass (upper curves)
averaged fractions of neutral gas in the universe. The observational
points for the volume averaged neutral fraction are from Bolton et
al.~(2007b), while the observed mass-fractions are from the damped
Ly$\alpha$ measurements of Prochaska et al.~(2005), and therefore
represent lower limits on the total HI content of the IGM. Both curves
show excellent agreement with these quantities, despite their
differing by 3 orders of magnitude.  In computing the volume averaged
neutral fraction we have followed standard practice and assumed
ionization equilibrium with an ionizing background at all
over densities. However in an IGM that includes dense regions that are
self-shielded, this value under estimates the true value. We note that
the inclusion of fully neutral gas at densities above the self
shielding over density does not modify the predicted value of effective
Ly$\alpha$ transmission, from which IGM properties are
inferred. However neutral hydrogen above the self-shielding threshold
does contribute significantly to the volume averaged neutral fraction
(in addition of course to the mass averaged neutral fraction)
interpreted from Ly$\alpha$ absorption spectra.

In the lower-left panel we plot the evolution of the ionizing photon
mean-free-path. The data points are based on Storrie-Lombardi et
al.~(1994). Again the model is in good agreement with the available
observations. We note that the observed mean-free-path is found from
the number density of Ly-limit systems and is independent of the
Ly$\alpha$ forest absorption derived quantities of ionization rate and
volume averaged neutral fraction, as well as being independent of the
HI mass-density measurements. Our simple model therefore
simultaneously reproduces the evolution of three independent measured
quantities. In the lower-right panel we plot the corresponding
evolution of the 21cm brightness temperature (dark lines). The grey
lines show the evolution of the filling factor of ionized regions
$(1-Q_{\rm i})$.

\section{Quasar Proximity Zones During Overlap}

\label{proximity}

\begin{figure*}
\includegraphics[width=15cm]{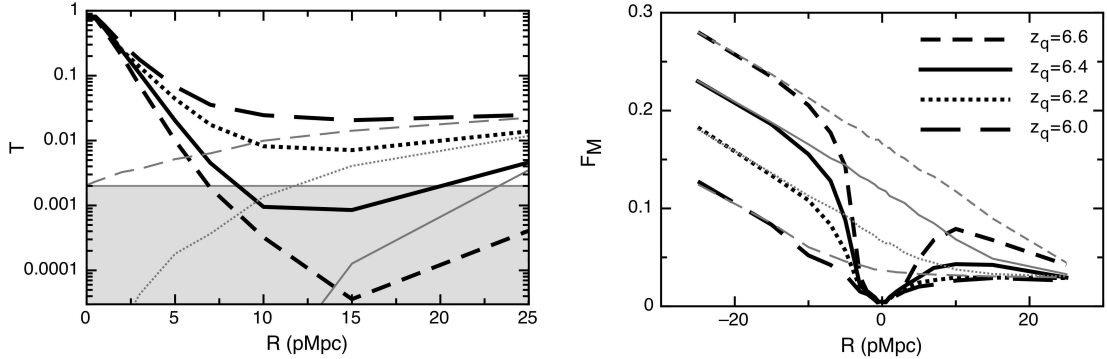} 
\caption{Impact of the quasar flux on local IGM properties. {\em Left:} Ly$\alpha$ transmission as a function of distance from the quasar in units of proper Mpc (dark curves). The shaded region shows the limit on transmission in the Gunn-Peterson troughs of the deepest available spectra. {\em Right:} Mass weighted neutral fraction as a function of distance from the quasar (dark curves). In each panel examples are shown corresponding to 4 different quasar redshifts. Also shown are the corresponding examples for the mean IGM, centered on the assumed quasar redshifts (grey curves).}
\label{fig2}
\end{figure*}

The enhanced ionization rate near quasars at moderate redshift
produces a region of thinned Ly$\alpha$ forest that extends for several Mpc
(e.g. Scott et al.~2000). This thinning of the Ly$\alpha$ forest is termed the
proximity effect and the region of enhanced ionization the proximity
zone. Prior to the end of reionization the proximity zone is not
defined since there is no ionizing background and instead the quasar
contributes to an enlarged, distinct HII region. In this section we
aim to model the effect of the quasar flux on the mass averaged
neutral fraction within the proximity zones of quasars during the
overlap era.

Our semi-analytic model provides a framework within which to model the
depletion of neutral hydrogen within the proximity zone. Since gas is neutral at over densities above $\Delta_{\rm i}$, we must estimate the change in $\Delta_{\rm i}$ induced by the quasar. We begin with the ionization
rate from our semi-analytic model as a function of radius from the
quasar host. We then add the ionization rate due to a quasar with
luminosity of $0.7\times10^{57}$s$^{-1}$, which corresponds to a quasar
with an absolute luminosity that is around a magnitude fainter than the most
luminous SDSS $z\sim6$ quasars. This process includes calculation of
biased reionization within the vicinity of the quasar host, but does
not include the increase in ionizing photon mean-free-path that would
result from the presence of quasar flux. We therefore underestimate
the ionization rate in the vicinity of the quasar. The most over dense regions of the IGM
are self shielding to ionizing radiation. The
over density of a clump at which gas becomes self shielding may be estimated from (Furlanetto \& Oh~2005; Bolton \& Haehnelt~2007b)
\begin{equation}
\label{DeltaSSC}
\Delta_{\rm SSC} = 50N\left(\frac{1+z}{7}\right)^{-3} \Gamma_{12}^{2/3},
\end{equation}
where we have neglected the mild dependence on temperature.
This expression assumes that the typical size of an absorber with over density $\Delta$ is the local Jeans length, and that the absorber becomes optically thick to Lyman limit photons when the column density $N_{\rm HI}$ exceeds $N\sigma_{\rm HI}^{-1}$, where $\sigma_{\rm HI}$ is the hydrogen photo-ionization cross-section at the Lyman limit. The coefficient $N$ has been previously assumed to equal unity, but is somewhat arbitrary, and we discuss its value below.

Our model for the reionization of the IGM surrounding a
quasar is not internally consistent, which would 
require a full numerical simulation. On
the other hand, such a simulation is currently beyond the available
numerical resolution over volumes sufficiently large to host a high
redshift quasar proximity zone, while the damped Ly$\alpha$ systems
thought to dominate the high redshift neutral gas are currently
subject to many model uncertainties (Nagamine et al.~2007). 
The inconsistency within our model can be traced partly to the fact that
$\Delta_{\rm i}$ is computed in a
non-equilibrium condition as reionization progresses, while
$\Delta_{\rm SSC}$ is computed in ionization equilibrium with an
estimate of the ionizing flux. However most
importantly, one has to assume a density profile in order to calculate a column
density. The density profiles which are assumed in the distribution
$P_{\rm V}(\Delta)$ differ from the top-hat density profile assumed in
calculation of $\Delta_{\rm SSC}$ (Furlanetto \& Oh~2005). In order
to make the model internally consistent we therefore choose the value
of $N$ in equation~(\ref{DeltaSSC}) at each redshift such that $\Delta_{\rm SSC}=\Delta_{\rm i}$ in
the mean ionizing background\footnote{The modification of the value $N$ in this process could be thought of as providing a correction that accounts for the effect of the density profile on the column density.}  (i.e. far from the quasar). The mass
averaged neutral fraction within the proximity zone at a radius $R$ is
then obtained from
\begin{equation}
F_{\rm M}(R)=\int_{0}^{\Delta_{\rm SSC}}d\Delta \Delta P_{\rm V}(\Delta) x_{\rm HI},
\end{equation}
where $x_{\rm HI}$ is the neutral fraction of hydrogen, which is evaluated assuming ionization equilibrium at over densities $\Delta<\Delta_{\rm SSC}$, and is equal to unity when $\Delta>\Delta_{\rm SSC}$.

An important ingredient in our modeling is the computation of the
local ionization rate at the retarded time along the line of
sight. The use of retarded time is particularly important for quasars
observed near the end of reionization since the ionization state of
the IGM changes dramatically during the light travel time of a quasar
proximity zone. In regions of the IGM observed in front of the quasar
(i.e. at lower redshift) we assume ionization equilibrium of the
hydrogen with the sum of quasar and ionizing back ground flux at all
radii. This is because the IGM is observed in photons that arrive at
the observer at the same time as photons emitted by the quasar (either
when observed in absorption against the quasar, or in 21cm
emission). However the 21cm emission from IGM behind the quasar can
only be subject to ionization by quasar flux if it is located
within a distance $R=c t_{\rm q}/2$ of the quasar (where
$t_{\rm q}$ is the quasar lifetime).

\begin{figure*}
\includegraphics[width=15.cm]{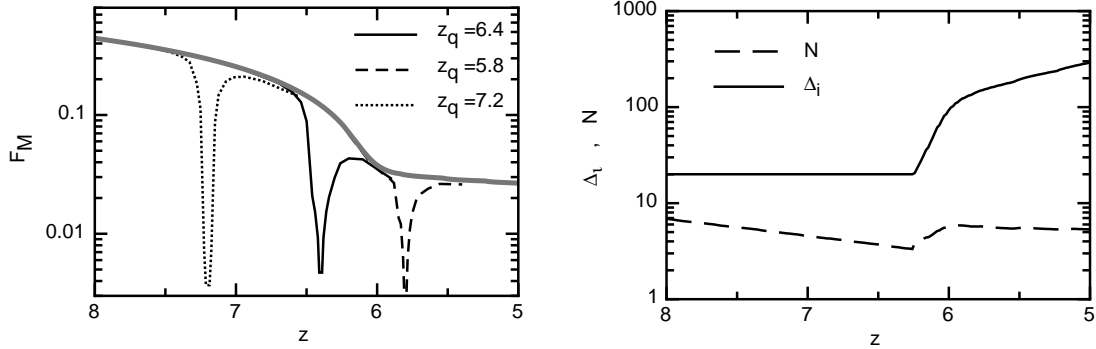} 
\caption{{\em Left:} Mass weighted neutral fraction as a function of redshift. The evolution for the mean IGM with redshift is shown as the grey line. The results of our modeling for the mass fraction in the vicinity of a quasar are shown as the dark curves for quasars at three different redshifts.  {\em Right:} The evolution of $\Delta_{\rm i}$ and $N$ as a function of redshift.   }
\label{fig3}
\end{figure*}

The resulting model proximity zones, described by the mass averaged
neutral fraction as a function of distance from the quasar (dark
curves) are plotted in the right-hand panel of
Figure~\ref{fig2}. Since the neutral gas in confined to discrete clumps following reionization, an individual line of sight through a proximity zone will not have a smooth profile in neutral hydrogen density. However our semi-analytic model is unable to compute individual realizations of the IGM, but rather computes the average behaviour which is represented by a smooth profile. Thus, examples of these smooth profiles are shown corresponding to 4
different quasar redshifts which cover the end of the overlap epoch (as
predicted within our model).  The curves show a proximity zone extending to
around 10 proper Mpc. The curves also show an asymmetry in the proximity
zone, which has a greater contrast behind the quasar. This asymmetry is
largest earlier in the overlap era. Also shown for comparison are the
corresponding examples for the mean IGM, centered on the assumed
quasar redshifts (grey curves).

We also compute the Ly$\alpha$ transmission as a function of distance
from the quasar, and plot the results in the left-hand panel of
Figure~\ref{fig2}. These transmission curves may be
compared with observations of high redshift quasars. The deepest
spectra of high redshift quasars reach a limit of $T\sim0.002$ towards
quasars at $z\sim6.3-6.4$. This limit is shaded grey. Our model
predicts that the Gunn-Peterson Trough will appear in the spectra of
quasars at $z\sim6.4$ (in agreement with observation), even though the
IGM is highly ionized. The model Gunn-Peterson trough
begins at a distance of approximately 7-8Mpc from the quasar and extends for
$\sim10$Mpc. The corresponding examples of transmission for the mean
IGM, centered on the assumed quasar redshifts are plotted for
comparison (grey curves). In this example the quasar has little impact
on the redshift at which the Gunn-Peterson trough would appear, since
the onset is at a large distance from the quasar.

In the left hand panel of Figure~\ref{fig3} we plot the mass weighted
neutral fraction as a function of redshift. The results of our
modeling for the mass fraction in the vicinity of a quasar are shown
as the dark curves for quasars at three different redshifts. For
comparison the evolution of the mean IGM with redshift is shown as the
grey line. On the right hand side we show the value of the pre factor
$N$ in equation~(\ref{DeltaSSC}) for $\Delta_{\rm SSC}$, as well as
$\Delta_{\rm i}$. Our model has $N\sim4$ for most of the redshift
range of interest, and shows that while it is not constant, $N$
evolves much more slowly than $\Delta_{\rm i}$.  Before proceeding, we
note that the quantitative predictions of our model will be sensitive
to the applicability of the assumed distribution of over densities $P_{\rm
V}(\Delta)$, which is not directly measured in numerical simulations
(Bolton \& Haehnelt~2007b).

\section{21cm Observations of Quasar Proximity Zones}

\label{21cmproximity}

\begin{figure*}
\includegraphics[width=15cm]{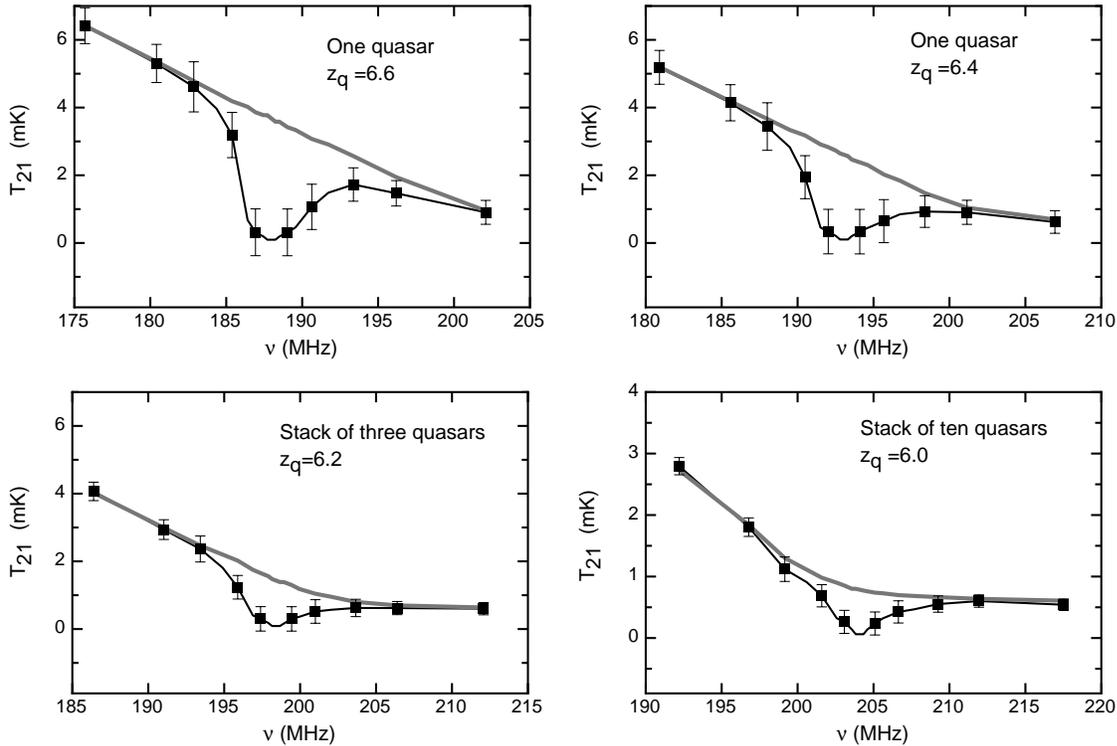} 
\caption{The 21cm brightness temperature as a function of observed frequency. The four panels show examples corresponding to 4 different quasar redshifts. Also shown are the corresponding examples for the mean IGM, centered on the assumed quasar redshifts (grey curves). At $z\geq6.4$, the error bars shown correspond to an observation of 1000 hours with the MWA using a maximum baseline corresponding to a 3.2' beam. At $z=6.2$ and $z=6.0$ the errors assume an average signal from stacks of 3 and 10 quasars respectively. }
\label{fig4}
\end{figure*}

We next estimate the 21cm signal corresponding to the proximity
zones described in the previous section. The 21cm brightness temperature contrast corresponding to IGM at the mean density is
\begin{equation}
T(R) = 22\mbox{mK}\left(\frac{1+z}{7.5}\right)^{1/2}(1-Q_{\rm i}F_{\rm M}(\Delta_{\rm i},R)).
\end{equation}
The resulting 21cm brightness
temperature profiles as a function of observed frequency are
shown in Figure~\ref{fig4} (dark lines). For a quasar at $z\sim6.6$
the model predicts that the expected contrast in front of the quasar
is only $\sim1$mK, while at redshifts beyond the quasar the contrast would
be as large as $5$mK. On the other hand, around a quasar at $z\sim6.0$
the contrast would only by $\sim0.5-1$mK. Also shown for comparison
are the corresponding examples for the mean IGM, centered on the
assumed quasar redshifts (grey curves).

We estimate the uncertainty for observations using the configuration
of the Murchison Widefield Array (MWA\footnote{see
http://www.haystack.mit.edu/ast/arrays/mwa/index.html}), which is
currently under construction and will comprise a phased array of 500
tiles (each tile will contain 16 cross-dipoles) distributed over an
area with diameter 1.5km.  The uncertainty was computed assuming 1000
hours of observing time. When forming a map from the available
visibilities it is assumed that resolution has been compromised for
lower noise in the image by choosing a maximum baseline to be included
(Geil \& Wyithe~2007). We chose a synthesised beam (full beam width at
half maximum) of $\theta_{\rm beam}=3.2'$ \footnote{This corresponds to 5.5' central peak to first null, which is often quoted as the resolution.} which would be appropriate
for quasar proximity zones. The thermal noise corresponding to this angular
scale for the MWA is (Geil \& Wyithe~2007)
\begin{equation} 
\Delta T = 17 \mbox{mK}\left(\frac{1+z}{8.65}\right)^{2.6}
\left(\frac{\Delta\nu}{0.07\mbox{MHz}}\right)^{-0.5}
\left(\frac{t_{\rm int}}{100\mbox{hr}}\right)^{-0.5}, 
\end{equation}
where $\Delta\nu$ is the width of the frequency bin and $t_{\rm int}$ is the
integration time.  At $z\geq6.4$, the error bars shown correspond to
an observation of a single quasar, while at $z=6.2$ and $z=6.0$ the
errors refer to the average signal from stacks of 3 and 10 quasars
respectively. These numbers are motivated by the expected number
counts in planned surveys and will be discussed in \S~\ref{counts}.
The sizes of the error bars correspond to binning over the interval in
between the points shown (so that the errors would be
independent). The proximity zones would be detectable with good
significance in the scenario described. 

We have computed line-of-sight 21cm spectra, while observations
will be made at finite resolution. In the case of spherical
proximity zones, finite resolution will introduce smoothing across
the boundaries in observed 21cm spectra. For this reason we have
restricted our analysis to spectra measured within a single
synthesised beam centered on the quasar line-of-sight. If the
proximity zone were spherical, the transverse size of the synthesised
beam ($\theta_{\rm beam}=3.2'$) at the edge of the proximity zone
would subtend $\sim20$ degrees, making the smoothing negligible.  Of
course the proximity zone is unlikely to be spherical. Suppose
that the quasar were beamed with an opening angle $\alpha$. If the
transverse extent of the synthesised beam were wider than the emission
region at the edge of the proximity zone, we would again expect
smoothing of the boundary in the observed 21cm spectrum. However,
given $\theta_{\rm beam}=3.2'$, the above argument implies that the
spectrum should not be subject to smoothing so long as $\alpha\ga20$
degrees (unless the quasar beaming is misaligned with the line of
sight by $\sim\alpha$).

Furthermore, in the examples shown at $z\leq6.2$ we have assumed that
the spectra from several quasars with the same luminosity and redshift
could be stacked. In practice this process would be subject to several
complications. First, the high redshift edge of the QSO proximity zone
may lie in the pre-overlap era. In this case we would expect
significant variation between quasars owing to the patchiness of
reionization. In addition, the separation in redshift of the quasars
available might exceed the depth of the proximity zones. Combining the
spectra of different quasars in order to increase signal-to-noise
would therefore be non-trivial.

Following the completion of overlap in a region of IGM the neutral gas
is in a collection of dense pockets rather than being diffusely
distributed in the IGM. There is therefore an additional component of
uncertainty for the spectra shown in Figure~\ref{fig4} owing to the
finite number of emission sources that contribute to the 21cm signal. Given a beam radius
$\theta_{\rm beam}$ and mass-averaged neutral fraction $F_{\rm M}$, there are
\begin{eqnarray}
\label{eq_N}
\nonumber
N_{\rm cl} &\sim& 400 \left(\frac{\theta_{\rm beam}}{3.2'}\right)^2 \left(\frac{F_{\rm M}}{0.03}\right)  \left(\frac{M_{\rm c}}{10^{9}M_\odot}\right)^{-1}  \\
&&\hspace{25mm}\times\left(\frac{\nu_{\rm bin}}{5\mbox{MHz}}\right) \left(\frac{1+z}{7}\right)^{0.9}
\end{eqnarray}
emission sources within a frequency bin of width $\nu_{\rm bin}$
assuming the neutral clumps to have a {\em baryonic} mass $M_{\rm c}$. Thus, we estimate that if the baryonic mass is limited to be the Jeans mass in an ionized IGM ($M_{\rm c}\sim10^9M_\odot$) then the component of uncertainty in a $\nu_{\rm bin}\sim5$MHz bin is around 5\%, while the existence of lower mass neutral clumps survived from the pre-reionization IGM would lead to an even smaller Poisson contribution. Of course a full calculation of $N_{\rm cl}$ would require a numerical simulation
to resolve the details of the damped Ly$\alpha$ and Ly-limit systems. However
equation~(\ref{eq_N}) suggests that the finite distribution of neutral
clumps will not contribute the dominant source of uncertainty in 21cm observations of quasar proximity zones with the MWA.

\section{The effect of Foreground Subtraction}

\label{fgproximity}

\begin{figure*}
\includegraphics[width=15cm]{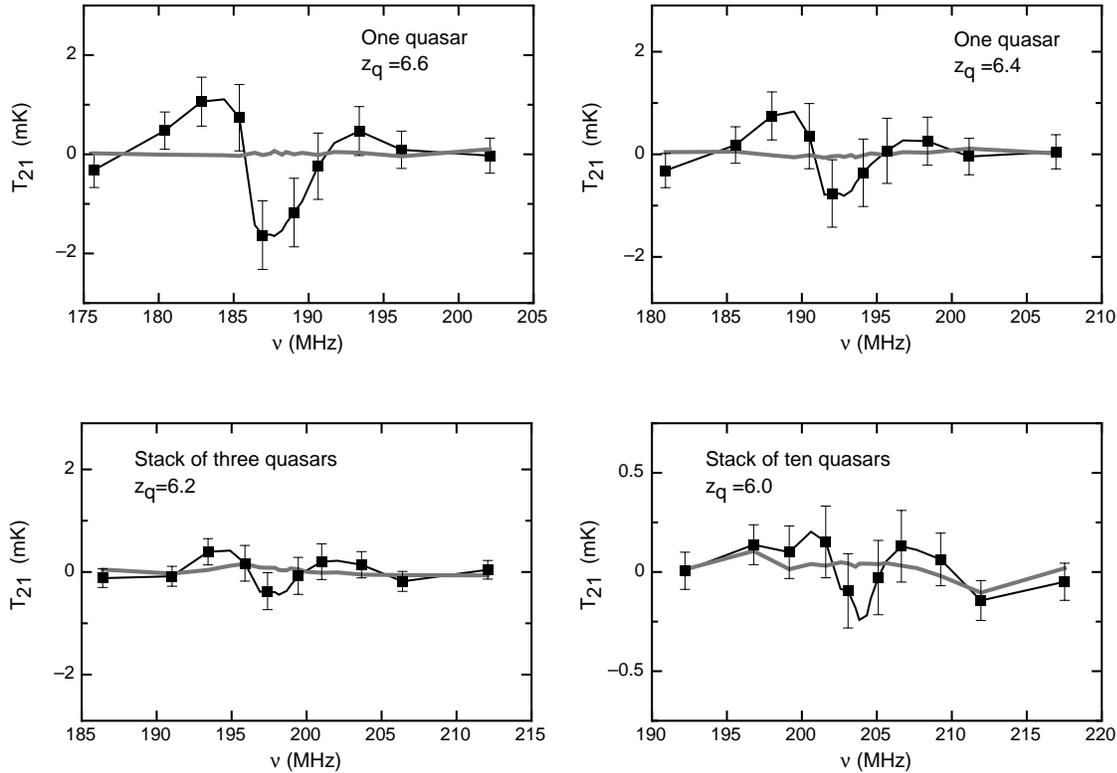} 
\caption{The 21cm brightness temperature as a function of observed frequency following subtraction of the best fit 4th order polynomial. The four panels show examples corresponding to the 4 different quasar redshifts described in Figure~\ref{fig4}. Also shown are the corresponding examples for the mean IGM, centered on the assumed quasar redshifts (grey curves). At $z\geq6.2$, the error bars shown correspond to an observation of 1000 hours with the MWA using a maximum baseline corresponding to a 3.2' beam. At $z=6.2$ and $z=6.0$ the errors assume an average signal from stacks of 3 and 10 quasars respectively.}
\label{fig5}
\end{figure*}

The detection of a proximity zone will require the subtraction of a
large foreground contribution to the redshifted 21cm signal. This
foreground is thought to be dominated by synchrotron radiation, both
galactic and extra galactic, and to be spectrally smooth (Oh \&
Mack~2003; Di~Matteo et al.~2002). The processed band-pass of the MWA
will be $\Delta\nu=32$MHz, which corresponds to a physical length of $\sim50$Mpc
at $z\sim6$. Thus the band pass that will be available to the MWA is similar in length to the profiles presented in
Figure~\ref{fig4}. The process of foreground
subtraction will render any line-of-sight fluctuations with wavelengths
comparable to, or larger than the band-pass undetectable. Thus, we
expect to lose the overall trend of the emission with redshift across
the band pass.

We take a simple approach to estimate the impact of foreground
removal, fitting and subtracting a fourth order polynomial from the
model spectra in Figure~\ref{fig4}. The resulting profiles are shown
in Figure~\ref{fig5} (dark lines). We see that the subtraction removes
the low order fluctuations, including the overall rise in intensity
across the band-pass. However the subtraction leaves fluctuations
around the slow rise due to the proximity zone. These fluctuations
will be detectable, and their amplitude will yield the mass-averaged
neutral fraction of hydrogen in the IGM at redshifts near that of the
quasar. The asymmetry of the HII region, which is clear in
Figure~\ref{fig4} would be difficult to detect in the foreground
removed spectra shown in Figure~\ref{fig5}. We will quantify this
statement below.

In Figure~\ref{fig5} we also show the corresponding
21cm spectra for the mean IGM centered on the assumed quasar redshifts,
with a fourth order polynomial fit subtracted as before. The resulting
profiles are shown as the grey curves. These profiles have no
detectable fluctuations. This is because the overlap epoch will take
place over a range of redshifts that is larger than the frequency
bandpass of the MWA. We therefore find that the global step will
be undetectable by the MWA (or any instrument with a comparable
bandpass). Note in this context that the modeling presented yields a global step
that is as rapid as allowed in a standard cosmological scenario.

\begin{figure*}
\includegraphics[width=15cm]{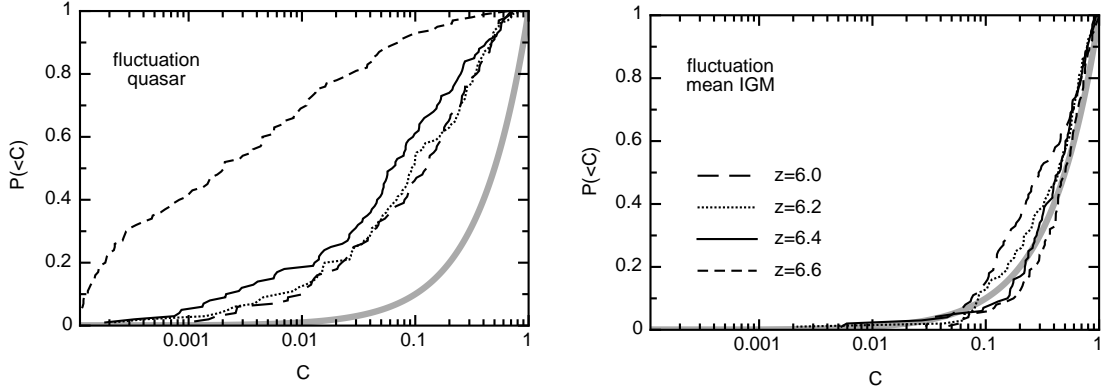} 
\caption{{\em Left panel:} Monte-Carlo realisations of the confidence with which the proximity zone can be distinguished from a smooth (modeled as a 4th order polynomial) evolution in the 21cm signal. Specifically, the quantity $C$ is the probability that a value of $\chi^2$ larger than that observed would arise by chance given a null-hypothesis consisting of the best fit 4th order polynomial. {\em Right panel:}  The corresponding distributions for spectra of the mean IGM. In each case the plot shows the cumulative distribution of $C$, at 4 redshifts corresponding to the examples shown in Figures~\ref{fig4}-\ref{fig5}. Also shown for comparison is the cumulative probability distribution $P_{\rm lin}(<C)=C$ (thick grey lines).}
\label{fig6}
\end{figure*}

We next quantify the significance with which
the quasar proximity zones could be detected in redshifted 21cm
spectra.  Our discussion is limited to determination of the confidence
with which the spectral feature at the redshift of the quasar (and its asymmetry, see
Figure~\ref{fig4}) could be detected. We define detection as the
statistically significant rejection of a null-hypothesis comprised of
the best fit 4th order polynomial to the spectra shown in
Figure~\ref{fig4}.  Our approach is to compute a set of Monte-Carlo
realisations of the 21cm spectrum by adding Gaussian noise to the
model spectra shown in Figure~\ref{fig4}. The Gaussian noise is
assumed to have a distribution of variance equal to the noise shown in
Figures~\ref{fig4}-\ref{fig5}. Using this set of noisy model spectra
we construct the cumulative distribution of the confidence with which
the HII region can be distinguished from a smooth (4th order
polynomial) evolution in the 21cm signal.  Specifically, we construct
the cumulative distribution of the confidence $C$, defined as the
probability that a value of $\chi^2$ larger than observed would arise
by chance given the null-hypothesis consisting of the best fit 4th
order polynomial. The resulting distributions are plotted in the left
hand panel of Figure~\ref{fig6}, at four redshifts corresponding to
the examples shown in Figures~\ref{fig4}-\ref{fig5}. As before we
assume a noise level corresponding to a single quasar at $z\geq6.4$,
and stacks of 3 and 10 quasars at $z=6.2$ and $z=6$ respectively. Note
that we have fitted spectra comprised of 10 points with a 5
parameter polynomial fit, leaving only 5 degrees of freedom. As a
result any residuals left following subtraction of the fit
lead to a high confidence for rejection of the null hypothesis. In
these examples, the spectral dip could be detected with 90\%
confidence in 50\% of cases for $6.0\la z\la6.4$. At $z\sim6.6$ the
spectral feature would be detected with 99\% confidence in more than 50\% of
cases. 

In the right hand panel of Figure~\ref{fig6} we show the corresponding
cumulative probability distributions for the confidence $C$ of
detecting a departure of the mean IGM spectrum from the 4th order
polynomial best fit. In each case we show cumulative distributions for
$C$ computed from mean IGM spectra centered at four redshifts
corresponding to the quasars in Figures~\ref{fig4}-\ref{fig5}. The
probability of the observed spectrum being inconsistent (in a $\chi^2$
sense) with the best fit 4th order polynomial is significantly less
for the mean IGM spectrum than for the quasar spectra, indicating that
the mean IGM signal is better modeled by a 4th order
polynomial than a quasar near-zone.

\begin{figure*}
\includegraphics[width=15cm]{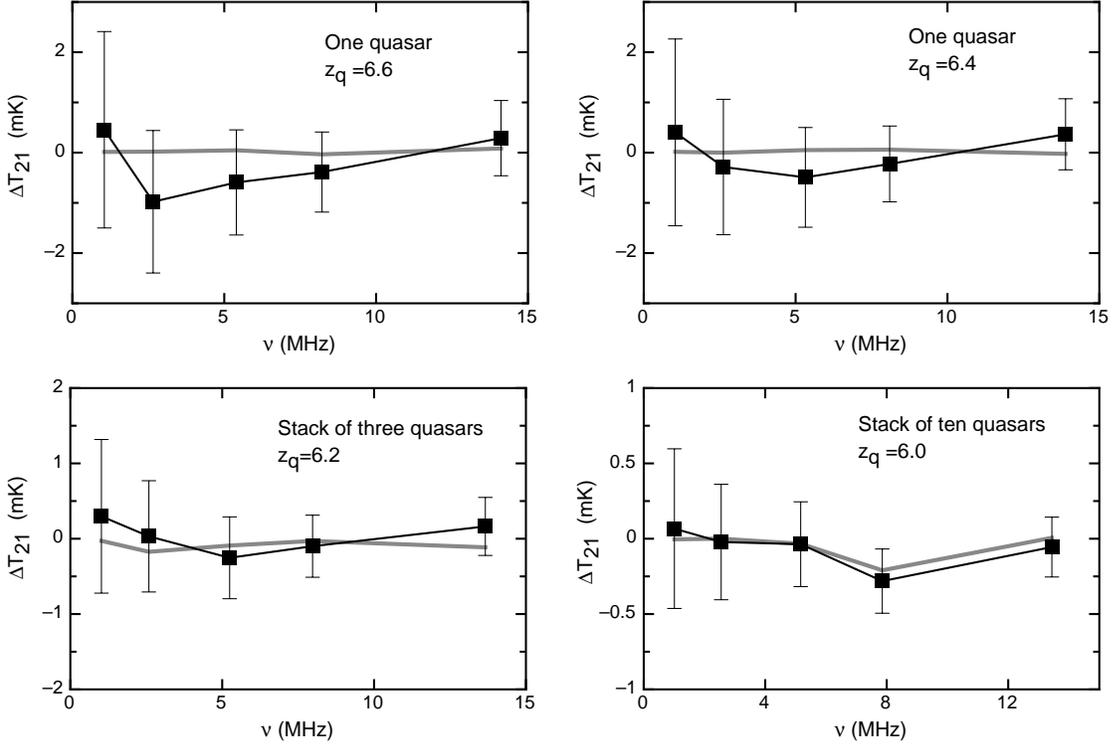} 
\caption{The difference ($\Delta T_{21}$) between the 21cm brightness temperature in front and behind the quasar following subtraction of the best fit 4th order polynomial. $\Delta T_{21}$ is plotted as a function of frequency difference relative to the quasar. A value of $\Delta T_{21}=0$ indicates a foreground removed spectrum that is symmetric about the quasar redshift. The four panels show examples corresponding to the 4 different quasar redshifts described in Figure~\ref{fig4}. Also shown are the corresponding examples for the mean IGM, centered on the assumed quasar redshifts (grey curves). At $z\geq6.2$, the error bars shown correspond to an observation of 1000 hours with the MWA using a maximum baseline corresponding to a 3.2' beam. At $z=6.2$ and $z=6.0$ the errors assume the average signal from stacks of 3 and 10 quasars respectively.}
\label{fig7}
\end{figure*}

For comparison, we also plot the cumulative distribution $P_{\rm
lin}(<C)=C$ in both panels of Figure~\ref{fig6}  (thick grey lines). In
cases where the 4th order polynomial perfectly models the spectrum, we
would expect no deviation of $P(<C)$ from $P_{\rm
lin}(<C)$. Figure~\ref{fig6} therefore illustrates that the probability of a
large value of $\chi^2$ is well in excess of random for spectra of quasar near zones 
at all redshifts considered. On the other hand, for spectra of the mean IGM, we find that the three higher redshifts
considered ($z\geq6.2$) have distributions $P(<C)$ that are nearly
indistinguishable from $P_{\rm lin}$, as might be expected from the
very smooth mean IGM spectra shown in Figure~\ref{fig4}. At $z=6.0$
the 4th order polynomial does not adequately describe the spectrum at
the end of overlap. As a result, $P(<C)>P_{\rm
lin}(<C)$. None-the-less, as mentioned above, $P(<C)$ is larger for
the quasar spectrum than for the mean IGM spectrum for each of the
four cases considered.

\begin{figure*}
\includegraphics[width=15cm]{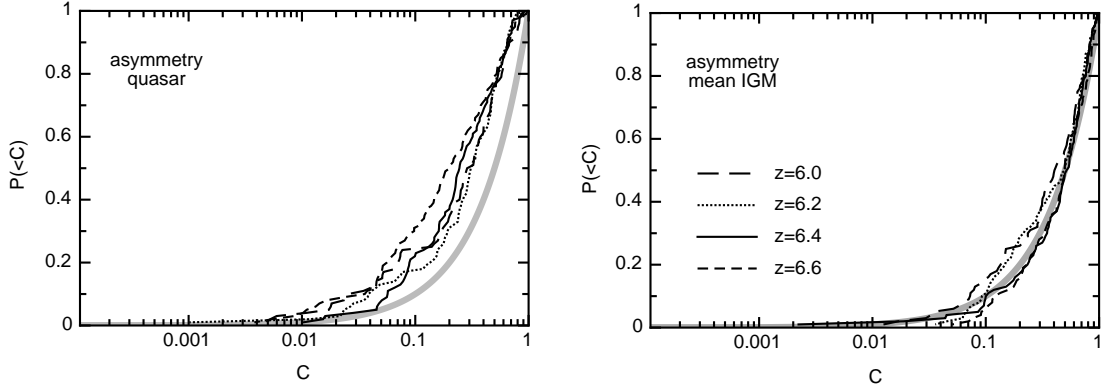} 
\caption{{\em Left panel:} Monte-Carlo realisations of the confidence with which asymmetry can be detected in the best fit foreground subtracted spectrum of a proximity zone. Specifically, the quantity $C$ is the probability that a value of $\chi^2$ larger than that observed would arise by chance given a null-hypothesis consisting of a symmetric proximity zone. {\em Right panel:}  The corresponding distributions for spectra of the mean IGM. In each case the plot shows the cumulative distribution of $C$, at 4 redshifts corresponding to the examples shown in Figures~\ref{fig4}-\ref{fig5}. Also shown for comparison is the cumulative probability distribution $P_{\rm lin}(<C)=C$ (thick grey lines).}
\label{fig8}
\end{figure*}

As noted earlier, the progression of overlap during the light travel
time across the proximity zone leads to the asymmetric 21cm spectra
shown in Figure~\ref{fig4}. However following the subtraction of a
4th-order polynomial, the asymmetry is less pronounced. To quantify
whether the asymmetry could be detected in the foreground removed
spectra, we subtract the points in the foreground removed spectra at
negative $R$ from points at positive $R$. This produces curves of residual
asymmetry $\Delta T_{21}(R)= T_{21}(R)-T_{\rm 21}(-R)$ consisting of 5
points which are shown in Figure~\ref{fig7}. Inspection of these
residuals indicates that foreground removal will prohibit the detection
of asymmetry, which would show up in Figure~\ref{fig7} as values of
$\Delta T_{21}$ that differ from zero. To quantify this statement, we
compute the value of $\chi^2$ relative to the null-hypothesis of a
symmetric model (which would equal 0 at each $R$).  We compute this
$\chi^2$ and the associated confidence $C$ for each of the Monte-Carlo
model spectra assuming a $\chi^2$ distribution with 5-2=3 degrees of
freedom (corresponding to 5 points with comparison to a straight
line). We then construct the cumulative probability distribution
$P(<C)$ as before. The resulting distributions are plotted in
Figure~\ref{fig8}, along with the cumulative probability distribution
$P_{\rm lin}(<R)=C$ (thick grey lines). The left panel shows results
for spectra of quasar proximity zones, while the right panel shows
results for spectra of the mean IGM. The distributions indicate that
asymmetry in the 21cm spectra of quasar near-zones could not be
detected with high confidence under the observational conditions assumed in this paper.

\section{Luminosity Function and Number Counts of High Redshift
Quasars}

\label{LF}

The advent of large multi-wavelength optical surveys in recent years
has allowed the detailed study of the quasar luminosity function to be
extended from redshifts corresponding to the peak of quasar
activity ($z\sim3$), out to the end of the reionization era at
$z\ga6$.  Firstly, the density of quasars at a redshift of
$z\sim6$ has been measured using the $8000$ square degrees of imaging from the Sloan Digital Sky Survey (SDSS), yielding 21 quasars with $z>5.8$ and
brighter than a $z$-band apparent AB-magnitude of $m_z=20$ (Fan et al.~2001; Fan et al.~2004; Fan et
al.~2006). In addition, fainter $z\sim6$ quasars have been discovered in a deeper survey of the SDSS equatorial stripe (yielding
5 quasars over $\sim125$ square degrees brighter than $m_z=21$; Jaing et
al.~2007). The combination of deep and wide surveys has allowed the
slope of the quasar luminosity function to be measured with high accuracy,
and yields a space density of quasars at $z\sim6$ which may be parameterised using the form
\begin{equation} 
\Theta_6(M_{1450}) = \Theta^*_6
10^{-0.4(\beta+1)(M_{1450}+26)}, 
\end{equation} 
where
$\Theta^*_6=(5.2\pm1.9)\times10^{-9}$Mpc$^{-3}$mag$^{-1}$, and
$\beta=-3.1\pm0.4$ (Jiang et al.~2007). 
The corresponding integral version of the luminosity function is 
\begin{eqnarray} 
\nonumber \Psi_6(M_{1450}) &=&
\int^{M_{1450}}_{-\infty}\Theta_6(M_{1450}) dM_{1450}\\ 
%\nonumber
%&=&\frac{\Theta(M_{1450})}{-0.4\ln(10)(\beta+1)}\\
&=&\Psi_6^*10^{-0.4(\beta+1)(M_{1450}+26)}, 
\end{eqnarray} 
where
\begin{equation} 
\Psi^*_6\equiv\frac{\Theta^*_6}{-0.4\ln(10)(\beta+1)}.
\end{equation} 
Comparison of the space density of luminous quasars at
$z\sim6$ (Fan et al.~2004) with the density measured at $z\sim4.3$ 
(Fan et al.~2001) shows an exponential decline in quasar number density with with redshift
\begin{equation}
\label{zev} 
\Psi(M_{1450}<-26.7,z) \propto 10^{B\times z}, 
\end{equation} 
where $B=-0.49\pm0.07$ (Wyithe \& Padmanabhan~2005).

The slope of the luminosity function ($\beta$) changes between
$z\sim4.3$ and $z\sim6$, becoming steeper towards high redshift. However we
will assume that $\beta$ is constant at $z\ga6$, noting that since we
are interested in extrapolation to quasars of lower luminosity than those already known, this 
will lead to conservative number counts. With the
assumption of constant $\beta$, and following equation~(\ref{zev}) we
next write 
\begin{equation} 
\Theta^*(z)=\Theta^*_6\times10^{B(z-6)},
\end{equation} 
yielding the differential and cumulative luminosity
functions 
\begin{equation}
\Theta(M_{1450},z)=\Theta_6^*\times10^{B(z-6)}10^{-0.4(\beta+1)(M_{1450}+26)},
\end{equation} 
and 
\begin{equation}
\Psi(M_{1450},z)=\Psi_6^*\times10^{B(z-6)}10^{-0.4(\beta+1)(M_{1450}+26)}.
\end{equation} 
These estimates provide a strong
empirical basis with which to predict the number counts of quasars at
moderately larger redshifts, but with luminosities comparable to those
currently observed.

\begin{figure*}
\includegraphics[width=15cm]{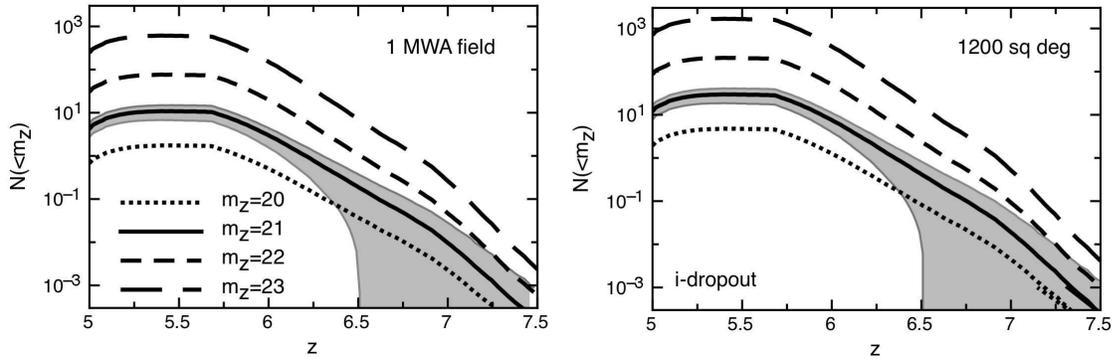} 
\caption{The number of quasars brighter than $m_z$ within a 32MHz band whose high frequency end observes the 21cm line at redshift $z$. We plot estimates of the 1-sigma range for a limit $m_z<21$. The left-hand and right-hand panels show the numbers for an MWA field, and a 1200 square degree area.}
\label{fig9}
\end{figure*}

\section{Estimated Number Counts for Future Surveys}

\label{counts}

Our goal in this section is to estimate the number of quasar proximity zones that might be available for study with low frequency arrays. As a specific example we consider the MWA, and multiply the density of quasars by the volume within an MWA
observation in order to estimate the potential number counts. Using relations from Furlanetto, Oh \& Briggs~(2006), we obtain the co-moving
volume of a cylinder of angular radius $\theta$ and depth
$\Delta\nu$ around a frequency $\nu$ corresponding to the 21cm line at
redshift $z$
\begin{equation}
V(\theta,\Delta\nu,z) = 4\times10^9\left(\frac{\theta}{12\mbox{deg}}\right)^{2}\left(\frac{1+z}{7}\right)^{0.9}\left(\frac{\Delta\nu}{32\mbox{MHz}}\right)\mbox{Mpc}^3.
\end{equation}
Combining this volume with the quasar luminosity function we obtain the number counts of quasars per frequency interval $\Delta\nu$ at a redshift $z$ brighter than $M_{1450}$
\begin{eqnarray}
\nonumber
\frac{dN(M_{1450},z)}{d\nu}\Delta\nu &=& V(\theta,\Delta\nu,z)\Psi(<M_{1450},z)\\
\nonumber
 &=& \frac{-\Theta_6^*\times10^{10}}{\ln(10)(1+\beta)}10^{-0.4(\beta+1)(M_{1450}+26)} \\
&\times& \left(\frac{1+z}{7}\right)^{1.9}\left(\frac{\Delta\nu}{32\mbox{MHz}}\right)10^{B(z-6)}.
\end{eqnarray}
Here we have noted that the field of view for the MWA is
\begin{equation}
\pi\theta^2\sim\frac{\lambda^2}{(4\mbox{m})^2}\times\frac{41253}{4\pi}\,\,\mbox{square degrees},
\end{equation}
with $\nu \lambda = c$ (yielding $\theta=12[(1+z)/7]$
degrees). Integrating over the redshift interval corresponding to the
MWA bandpass of $\Delta\nu=32$MHz, we find
\begin{equation}
\label{MWAcounts}
N(M_{1450},z) = \int_{\nu-\Delta\nu}^\nu \frac{d\nu}{\Delta\nu} \frac{dN(<M_{1450},z)}{d\nu}\Delta\nu, 
\end{equation}
where the redshift $z=7(\nu/204.1\mbox{MHz})-1$ corresponds to the
21cm line redshifted to the high frequency end of the bandpass.  We
next convert these number counts for an absolute AB-magnitude
$M_{1450}$ to observed number counts for an apparent magnitude limit
using the median spectrum from the LBQS (Francis et al.~1991), as well
as the SDSS transmission curves for the $i$ and $z$-filters. We assume
that flux is fully absorbed in the Ly$\alpha$ forest below an
observed wavelength $1216(1+z)\AA$, which is appropriate for $z\ga5.8$
quasars (e.g. White, Becker, Fan, Strauss~2003).

We present number counts for one MWA field, and also
for a 1200 square degree region.  These are shown in Figure~\ref{fig9}
as a function of redshift for four different $z$-band limiting
AB-magnitudes. The counts are presented assuming
$\Delta\nu=32$MHz bandpass (which extends over a large fraction of a
redshift unit). In Figure~\ref{fig9} the labeled redshift corresponds
to the frequency of the redshifted 21cm line at the high frequency end
of the observing band (see equation~\ref{MWAcounts}). As a concrete
example, it is anticipated that the forthcoming Sky Mapper survey will
reach $i\sim23$ in its six-epoch survey (Keller et al.~2007). Since
$z\sim6$ quasars are discovered using an $i$-dropout criteria
($i-z>2.2$; Fan et al.~2001) the corresponding limiting magnitude
achieved would be $m_z\sim21$. This limit is shown as the solid line
in Figure~\ref{fig9}, along with the grey shaded region which refers
to the 1-sigma uncertainty in the observed luminosity function. The
$i$-dropout requirement limits discovery to $z\ga5.75$, which produces
the plateau at lower redshifts. Our estimates suggest that at most one
$z>6.5$ quasar (1-$\sigma$ upper limit) will be discovered in a 1200
square degree region at $m_z<21$, while one $z>6.25$ quasar would be
found in each MWA field. At slightly lower redshifts the number counts
will allow for stacking of signal from a number of quasars. For
example, we would expect to find $\sim5$ quasars with $z>6$ per MWA
field.

\section{Conclusion}

\label{conclusion}

The introduction of low-frequency radio arrays over the coming decade
is expected to revolutionize the study of the reionization
epoch. Several studies have been published previously, arguing that
the observation of the contrast in redshifted 21cm emission between a
large HII region and the surrounding neutral IGM will be the simplest
and most easily interpreted signature. These studies have focused on
the detection of an HII region, formed by the ionizing flux from a
quasar generating an ionized bubble in a significantly neutral
IGM. However the highest redshift quasars so far discovered (at
$z\sim6.4$) suggest that the IGM along those lines-of-sight is
substantially ionized. Thus, more distant quasars would need to be
discovered in order to find an HII region surrounding a previously
known source. However quasars are observed to be extremely rare at
high redshifts, and as discussed in \S~\ref{counts}, the prospects for their
discovery at $z>6.5$ in a region of around 1000 square degrees are not
good. Alternatively the HII region might be found directly from 21cm
data. While first generation instruments will detect an individual HII
region at modest signal to noise, it has been suggested that a matched
filter approach could be used to find HII regions in a blind search
(Datta, Bharadwaj \& Choudhury~2007). On the other hand, quasar HII
regions will have a very complex geometry during the overlap epoch,
even if they emit isotropically (Geil \& Wyithe~2007), making the
blind detection via a matched filter approach more difficult.

In this paper we have investigated the prospects for detection of
proximity zones in the highly ionized IGM surrounding quasars during
the late stages of the overlap era. We have concentrated on quasars at
redshifts where they are already known to exist in sufficient numbers
to make the measurements practical.  We employ a semi-analytic model
which reproduces several post overlap properties of the IGM, including
the ionizing photon mean-free-path, the mass-averaged density of
neutral gas and the hydrogen ionization rate. In agreement with more
sophisticated numerical simulations (e.g. Lidz et al.~2007; Bolton \&
Haehnelt~2007a), this model predicts that the Gunn-Peterson Trough will
appear in the spectra of quasars at $z\sim6.4$, even though the IGM is
highly ionized by that time.  We show that while quasars at $z<6.5$
are likely observed in the post overlap IGM, they will still provide
valuable probes of the reionization era. This usefulness arises
firstly because dense pockets of neutral gas will continue to provide a 21cm
signal even in a highly ionized IGM. Secondly, the light-travel time
across a quasar proximity zone is probably comparable to the duration
of hydrogen overlap. As a result, while the IGM studied in Ly$\alpha$ absorption
along the line-of-sight to the quasar may be highly ionized, the IGM
observed "behind" the quasar would be in an earlier stage of overlap
and so more neutral.

We have estimated the 21cm signal corresponding to quasar proximity zones
as a function of distance from the quasar. At $z\sim6.4$ our model
predicts that while the expected contrast in front of the quasar is less than
$\sim1$mK, at redshifts beyond the quasar the contrast would be as
large as $5$mK. On the other hand, around a quasar at $z\sim6.0$ the
contrast would be only $\sim0.5-1$mK. Assuming observations using the
configuration of the MWA, with 1000 hours of observing time and a
maximum baseline corresponding to a 3.2' beam we find that these
contrasts could be detected in observations of individual quasars at $z\ga6.4$,
while at $z\leq6.2$ detection would require a stack of observations
for several quasars.

In practice he detection of a proximity zone will require the subtraction of the
large foreground component which dominates the redshifted 21cm
signal. The process of foreground subtraction will render any
line-of-sight fluctuations with wavelengths comparable to, or larger
than the band-pass undetectable. We therefore fit and subtract a
fourth order polynomial to our model proximity zone spectra. We find
that the subtraction removes the low order fluctuations including the
overall rise across the band-pass.  However the
subtraction leaves residual fluctuations due to the proximity zone
that would be detectable with the MWA, although the asymmetry of the
proximity zone due to the evolution of the IGM will not.  In contrast
we find that foreground subtraction from the 21cm emission spectra
corresponding to the mean IGM leaves no detectable fluctuations. This
is because the overlap epoch will take place over a range of redshift
that is larger than the frequency bandpass of the MWA. Foreground
subtraction will render the global step in 21cm emission undetectable
within a single $\Delta\nu=32$MHz bandpass.

In our model we have used an analytic probability distribution for the
over density which was fit to numerical simulations (Miralda-Escude et
al.~2000). This model allows reionization to be computed in an
inhomogeneous IGM, and provides a framework within which to model the
progression of reionization from the era prior to overlap when the
neutral gas is found predominantly in the mean IGM to post overlap
where dense systems (DLAs) dominate the neutral hydrogen content of
the universe (Prochaska et al.~2005). At redshifts lower than
considered in this paper ($z\la5$) observations of proximate DLAs
(those within 3000 km/s of the observed quasar) show a density of
neutral hydrogen that is comparable to the mean IGM (Prochaska et
al.~2007). On the other hand, given the measured galaxy bias of DLAs,
an excess of DLAs would be expected in the over dense IGM where
the proximate DLAs are observed. Prochaska et al.~(2007) interpret
this result as evidence that the quasar flux creates a proximity zone
in the DLAs. Our model includes the mean enhancement of over density in
the IGM surrounding the quasar but does not include a galaxy bias for
the galaxies that are thought to host the DLAs at more intermediate
redshifts. Thus, we expect the model to provide a qualitatively
correct description prior to and during reionization. However at the
lowest redshifts considered the unknown properties of DLAs could
modify our model predictions.

The density of quasars at $z\sim6$ is now well constrained (Fan et
al.~2006). We have employed the latest measurements of quasar
densities at high redshift to estimate the number of quasars that will
be discovered in optical--near-IR surveys, with specific reference to
the numbers that may be found in MWA fields. Assuming that the MWA
fields can be aligned with such surveys we estimate the number of
quasars that will be discovered per MWA field. One particular example
is the Sky Mapper survey (Keller et al.~2007), which will find around
one quasar at $z>6.25$, and around $5$ quasars at $z>6$ per MWA
field. Surveys for high redshift quasars (e.g. Sky Mapper) will cover a
much larger fraction of the sky than is planned for redshifted 21cm
observations. The fact that upcoming surveys will find more that 1
$z>6$ quasar per MWA field is therefore important, since it means that
redshifted 21cm observations do not need to be made in fields where
quasars have been previously discovered.

In summary we find that if 21cm foregrounds can be subtracted to a
level below the thermal noise the 21cm emission, then proximity zones
around high redshift quasars will provide a probe of the very end of
the overlap era. These 21cm proximity zones will provide a bridge
between measurements of 21cm intensity fluctuations during the peak of
the reionization era and studies of Ly$\alpha$ absorption following the
completion of reionization, and so facilitate study of the entire evolution
of the ionization state of the IGM.

\bigskip

{\bf Acknowledgments} The research was supported by the Australian Research
Council. The author acknowledges helpful conversations with Abraham Loeb,  Paul Geil and James Bolton, and would like to thank an anonymous referee for suggestions which have improved this paper.

\newcommand{\noopsort}[1]{}

\label{lastpage}

\begin{thebibliography}{}

\bibitem[]{}
Barkana, R., Loeb, A. 2001, {Phys. Rep.}, {349}, 125

\bibitem[Barkana \& Loeb(2005)]{2005MNRAS.363L..36B} Barkana, R., \& Loeb, 
A.\ 2005, \mnras, 363, L36 

\bibitem[Barkana \& Loeb(2005)]{2005ApJ...626....1B} Barkana, R., \& Loeb, 
A.\ 2005, \apj, 626, 1 

\bibitem[Barkana \& Loeb(2005)]{2005ApJ...624L..65B} Barkana, R., \& Loeb, 
A.\ 2005, \apjl, 624, L65 

\bibitem[Becker et al.(2007)]{2007ApJ...662...72B} Becker, G.~D., Rauch, M., \& Sargent, W.~L.~W.\ 2007, \apj, 662, 72 

\bibitem[Bolton \& Haehnelt(2007a)]{2007MNRAS.374..493B} Bolton, J.~S., \& 
Haehnelt, M.~G.\ 2007a, \mnras, 374, 493 

\bibitem[Bolton \& Haehnelt(2007b)]{} Bolton, J.~S., \& 
Haehnelt,
M.~G.\ 2007b, submitted to \mnras

\bibitem[Bond et al.(1991)]{1991ApJ...379..440B} Bond, J.~R., Cole, S., 
Efstathiou, G., \& Kaiser, N.\ 1991, \apj, 379, 440 

\bibitem[\protect\citeauthoryear{{Chen}, {Prochaska} \& {Gnedin}}{{Chen}
  et~al.}{2007}]{Chen2007}
{Chen} H.-W.,  {Prochaska} J.~X.,    {Gnedin} N.~Y.,  2007, \apjl, 667, L125

\bibitem[Datta et al.(2007)]{2007astro.ph..3677D} Datta, K.~K., Bharadwaj, S., \& Choudhury, T.~R.\ 2007, ArXiv Astrophysics e-prints, arXiv:astro-ph/0703677

\bibitem[]{} Dijkstra, M., Haiman, Z., Rees, M.~J., \& Weinberg, D.~H. 2004, 
\apj, { 601}, 666

\bibitem[Di Matteo et al.(2002)]{2002ApJ...564..576D} Di Matteo, T., Perna, R., Abel, T., \& Rees, M.~J.\ 2002, \apj, 564, 576

\bibitem[]{} Efstathiou, G. 1992, { Mon. Not. R. Astron. Soc.}, { 256},
43

\bibitem[Fan et al.(2001)]{2001AJ....121...54F} Fan, X., et al.\ 2001, \aj, 
121, 54 

\bibitem[Fan et al.(2004)]{2004AJ....128..515F} Fan, X., et al.\ 2004, \aj, 
128, 515 

\bibitem[Fan et al.(2006)]{2006AJ....132..117F} Fan, X., et al.\ 2006, \aj, 132, 117 

\bibitem[Francis et al.(1991)]{1991ApJ...373..465F} Francis, P.~J., Hewett, 
P.~C., Foltz, C.~B., Chaffee, F.~H., Weymann, R.~J., \& Morris, S.~L.\ 
1991, \apj, 373, 465 

\bibitem[Furlanetto et al.(2004)]{2004ApJ...613...16F} Furlanetto, S.~R., Zaldarriaga, M., \& Hernquist, L.\ 2004, \apj, 613, 16 

\bibitem[Furlanetto 
\& Oh(2005)]{2005MNRAS.363.1031F} Furlanetto, S.~R., \& Oh, S.~P.\ 2005, \mnras, 363, 1031 


\bibitem[Furlanetto et al.(2006)]{2006PhR...433..181F} Furlanetto, S.~R., 
Oh, S.~P., \& Briggs, F.~H.\ 2006, Phys. Rep., 433, 181 


\bibitem[Geil \& Wyithe(2007)]{2007arXiv0708.3716G} Geil, P.~M., \& Wyithe, 
S.\ 2007, ArXiv e-prints, 708, arXiv:0708.3716 


\bibitem[\protect\citeauthoryear{{Gnedin}}{{Gnedin}}{2007}]{Gnedin2007b}
{Gnedin} N.~Y.,  2007, ArXiv e-prints, 709

\bibitem[Gnedin \& Fan(2006)]{2006ApJ...648....1G} Gnedin, N.~Y., \& Fan, 
X.\ 2006, \apj, 648, 1 

\bibitem[\protect\citeauthoryear{{Gnedin}, {Kravtsov} \& {Chen}}{{Gnedin}
  et~al.}{2007}]{Gnedin2007a}
{Gnedin} N.~Y.,  {Kravtsov} A.~V.,    {Chen} H.-W.,  2007, ArXiv e-prints, 707

\bibitem[Gnedin \& Shaver(2004)]{2004ApJ...608..611G} Gnedin, N.~Y., \& Shaver, P.~A.\ 2004, \apj, 608, 611

\bibitem[Jiang et al.(2007)]{2007arXiv0708.2578J} Jiang, L., et al.\ 2007, 
ArXiv e-prints, 708, arXiv:0708.2578 

\bibitem[Keller et al.(2007)]{2007PASA...24....1K} Keller, S.~C., et al.\ 
2007, Publications of the Astronomical Society of Australia, 24, 1 

\bibitem[Kohler et al.(2005)]{2005ApJ...633..552K} Kohler, K., Gnedin, N.~Y., Miralda-Escud{\'e}, J., \& Shaver, P.~A.\ 2005, \apj, 633, 552

\bibitem[]{} 
Leitherer, C., et al.\ 1999, \apjs, 123, 3

\bibitem[Lidz et al.(2007)]{2007astro.ph..3667L} Lidz, A., McQuinn, M., Zaldarriaga, M., Hernquist, L., \& Dutta, S.\ 2007, ArXiv Astrophysics e-prints, arXiv:astro-ph/0703667 

\bibitem[Loeb \& Zaldarriaga(2004)]{2004PhRvL..92u1301L} Loeb, A., \& Zaldarriaga, M.\ 2004, Physical Review Letters, 92, 211301 

\bibitem[Mesinger et al.(2004)]{2004ApJ...613...23M} Mesinger, A., Haiman, Z., \& Cen, R.\ 2004, \apj, 613, 23 

\bibitem[Miralda-Escud{\'e} et al.(2000)]{2000ApJ...530....1M} Miralda-Escud{\'e}, J., Haehnelt, M., \& Rees, M.~J.\ 2000, \apj, 530, 1 

\bibitem[Nagamine et al.(2007)]{2007ApJ...660..945N} 
Nagamine, K., Wolfe, 
A.~M., Hernquist, L., \& Springel, V.\ 2007, \apj, 660, 945 

\bibitem[Oh \& Mack(2003)]{2003MNRAS.346..871O} Oh, S.~P., \& Mack, K.~J.\ 2003, \mnras, 346, 871

\bibitem[]{}
Press, W., Schechter, P. 1974, {ApJ.}, {187}, 425

\bibitem[Prochaska et al.(2005)]{2005ApJ...635..123P} Prochaska, J.~X., Herbert-Fort, S., \& Wolfe, A.~M.\ 2005, \apj, 635, 123 

\bibitem[Prochaska et al.(2007)]{} Prochaska, J.~X., Henawi, J.~F., Herbert-Fort, S.,\ 2007, astro-ph/0703594 

\bibitem[Scott et al.(2000)]{2000ApJS..130...67S} Scott, J., Bechtold, J., 
Dobrzycki, A., \& Kulkarni, V.~P.\ 2000, \apjs, 130, 67 


\bibitem[Shaver et al.(1999)]{1999A&A...345..380S} Shaver, P.~A., Windhorst, R.~A., Madau, P., \& de Bruyn, A.~G.\ 1999, A\&A, 345, 380

\bibitem[Spergel et al.(2007)]{2007ApJS..170..377S} Spergel, D.~N., et al.\ 2007, \apjs, 170, 377 

\bibitem[Srbinovsky \& Wyithe(2007)]{2007MNRAS.374..627S} Srbinovsky, 
J.~A., \& Wyithe, J.~S.~B.\ 2007, \mnras, 374, 627 

\bibitem[Srbinovsky \& Wyithe(2008)]{} Srbinovsky, 
J.~A., \& Wyithe, J.~S.~B.\ 2008, \mnras, Submitted 

\bibitem[Storrie-Lombardi et al.(1994)]{1994ApJ...427L..13S} 
Storrie-Lombardi, L.~J., McMahon, R.~G., Irwin, M.~J., \& Hazard, C.\ 1994, \apjl, 427, L13 

\bibitem[]{} Thoul, A.~A., \& Weinberg, D.~H.
1996, { Astrophys. J.}, {465}, 608

\bibitem[Tozzi et al.(2000)]{2000ApJ...528..597T} Tozzi, P., Madau, P., Meiksin, A., \& Rees, M.~J.\ 2000, \apj, 528, 597

\bibitem[]{}
White, R., Becker, R., Fan, X., Strauss, M. 2003, {Astron J.}, {126}, 1 

\bibitem[]{}
Wyithe, J.~S.~B, Bolton,~J.~S., Haehnelt,~M. 2007, {MNRAS}, {submitted}

\bibitem[]{}
Wyithe, J.~S.~B, Loeb, A. 2003, {ApJ}, {586}, 693

\bibitem[]{}
Wyithe, J.S.B, Loeb, A. 2004, {Nature}, {427}, 815

\bibitem[Wyithe \& Loeb(2004b)]{2004ApJ...610..117W} Wyithe, J.~S.~B., \& Loeb, A.\ 2004b, \apj, 610, 117 

\bibitem[Wyithe \& Loeb(2007)]{2007arXiv0708.3392W} Wyithe, S., \& Loeb, 
A.\ 2007, ArXiv e-prints, 708, arXiv:0708.3392 

\bibitem[]{} Wyithe, J.~S.~B., Loeb, A., Barnes, D.G., ~2005, \apj, 634, 715

\bibitem[]{}
Wyithe, J.S.B, Loeb, A., Carilli, C. 2005, \apj, 628, 575

\bibitem[Wyithe \& Padmanabhan(2006)]{2006MNRAS.366.1029W} Wyithe, 
J.~S.~B., \& Padmanabhan, T.\ 2006, \mnras, 366, 1029 


\end{thebibliography}
\end{document}